\documentclass[prl,aps,showpacs,reprint]{revtex4-1}
\usepackage{amsmath}
\usepackage{amssymb}
\usepackage{ntheorem}

\usepackage[T1]{fontenc}

\usepackage[bookmarks=false]{hyperref}
\hypersetup{colorlinks=true,citecolor=blue,linkcolor=red,urlcolor=blue,pdfstartview=FitH,bookmarksopen=true}

\DeclareMathOperator{\Tr}{Tr}
\DeclareMathOperator{\NOT}{NOT}
\newcommand{\bra}[1]{\langle #1\rvert}
\newcommand{\ket}[1]{\lvert #1\rangle}

\newcommand{\I}{\mathrm{I}}

\newtheorem*{theorem}{Theorem}

\begin{document}
\title{Measure-Independent Freezing of Quantum Coherence}
\author{Xiao-Dong Yu}
\affiliation{Department of Physics, Shandong University, Jinan 250100, China}
\author{Da-Jian Zhang}
\affiliation{Department of Physics, Shandong University, Jinan 250100, China}
\author{C. L. Liu}
\affiliation{Department of Physics, Shandong University, Jinan 250100, China}
\author{D. M. Tong}
\email{tdm@sdu.edu.cn}
\affiliation{Department of Physics, Shandong University, Jinan 250100, China}
\date{\today}
\begin{abstract}
We find that all measures of coherence are frozen for an initial state in a strictly incoherent channel if and only if the relative entropy of coherence is frozen for the state. Our finding reveals the existence of measure-independent freezing of coherence, and provides an entropy-based dynamical condition in which the coherence of an open quantum system is totally unaffected by noise.
\end{abstract}
\maketitle

Quantum coherence is a fundamental feature of quantum mechanics, describing the capability of a quantum state to exhibit quantum interference phenomena. The coherence effect of a state is usually ascribed to the off-diagonal elements of its density matrix with respect to a particular reference basis, which is determined according to the physical problem under consideration. It is an essential ingredient in quantum information processing \cite{Nielsen.Chuang2000}, and plays a central role in emergent fields, such as quantum metrology \cite{Giovannetti.etal2004,Demkowicz-Dobrzanski.Maccone2014,Giovannetti.etal2011}, nanoscale thermodynamics \cite{Aberg2014,Narasimhachar.Gour2015,Cwiklinski.etal2015,Lostaglio.etal2015a,Lostaglio.etal2015b,Vazquez.etal2012,Karlstroem.etal2011}, and quantum biology \cite{Plenio.Huelga2008,Rebentrost.etal2009,Lloyd2011,Li.etal2012,Huelga.Plenio2013}.

It is only recent years that the quantification of coherence has become a hot topic due to the development of quantum information science, although the theory of quantum coherence is historically well developed in quantum optics \cite{Glauber1963,Sudarshan1963,Mandel.Wolf1995}. A rigorous framework to quantify the coherence of quantum states in the resource theories has been recently proposed after a series of efforts \cite{Baumgratz.etal2014,Aberg2006,Levi.Mintert2014,Marvian.Spekkens2013,Girolami2014,Bartlett.etal2006,Gour.Spekkens2008,Vaccaro.etal2008,Toloui.etal2011,Marvian.Spekkens2014}. By following the rigorous framework comprising four postulates \cite{Baumgratz.etal2014}, a number of coherence measures based on various physical contexts have been put forward. The $l_1$ norm of coherence and the relative entropy of coherence were first suggested as two coherence measures based on distance. The coherence measures based on entanglement \cite{Streltsov.etal2015}, the coherence measures based on operation \cite{Winter.Yang2016,Chitambar.etal2016}, and the coherence measures based on convex-roof construction \cite{Yuan.etal2015,Du.etal2015b} were subsequently proposed. With coherence measures, various properties of quantum coherence, such as the relations between quantum coherence and other quantum resources \cite{Streltsov.etal2015,Yao.etal2015,Xi.etal2015}, the quantum coherence in infinite-dimensional systems \cite{Zhang.etal2016,Xu2016}, the complementarity relations of quantum coherence \cite{Cheng.Hall2015}, and the measure of macroscopic coherence \cite{Yadin.Vedral2016}, have been discussed.

Quantum coherence is a useful physical resource, but coherence of a quantum state is often destroyed by noise. A challenge in exploiting the resource is to protect coherence from the decoherence caused by noise, as the loss of coherence may weaken the abilities of a state to perform quantum information processing tasks. Today, after having been equipped with the knowledge of coherence measures, it becomes possible to analyze under which dynamical conditions the coherence of an open system is frozen in a noisy channel. Studies on this topic have been started in Ref. \cite{Bromley.etal2015}, where the authors found that the coherence measures based on bona fide distances are frozen for some initial states of a quantum system with even number of qubits undergoing local identical bit flip channels. This finding illustrates that there exist such quantum states of which some coherence measures remain constant in certain noisy channels, and hence the ability of such states to perform quantum information processing tasks is not weakened by the noise if the ability exploited in the task is based on these frozen coherence measures.

However, some coherence measures being frozen do not imply other coherence measures being frozen too, since different coherence measures result in different orderings of coherence in general \cite{Liu.etal2016}. Freezing of coherence is dependent on the coherence measures adopted in general. Although a noisy channel may not weaken some abilities of a quantum state if these abilities are based on the frozen coherence measures, it can still weaken the other abilities that are based on unfrozen coherence measures. Only the states with measure-independent freezing of coherence can keep all the abilities of coherence resource totally unaffected. Here, the phrase, measure-independent freezing of quantum coherence, means that coherence of some states is frozen independently of coherence measures, i.e., all coherence measures of the states are frozen in certain channels. The question then is: Under which dynamical conditions does the measure-independent freezing phenomenon occur for an open quantum system in a noisy channel? This is an important issue, since only in this case the coherence of an open system is totally unaffected by noise. In this letter, we address this issue.

To present our finding clearly, we need first to recapitulate some notions, such as incoherent states, incoherent operations, strictly incoherent operations, and coherence measures.

An incoherent state is defined as
\begin{equation}
\delta=\sum_ip_i\ket{i}\bra{i},
\label{eq:incoherentstates}
\end{equation}
where $\{\ket{i}\}$ represents a fixed reference basis, and $p_i$ are probabilities. The set of all incoherent states is denoted by $\mathcal{I}$. All other states which cannot be written as diagonal matrices in this basis are called coherent states. Hereafter, we use $\rho$ to represent a general state, and $\delta$ specially to denote an incoherent state.

An incoherent operation or an incoherent channel, i.e., an incoherent completely positive trace-preserving map (incoherent CPTP) map, is defined as
\begin{equation}
\Lambda(\rho)=\sum_nK_n\rho K_n^\dagger,
\label{eq:incoherentoperations}
\end{equation}
where the Kraus operators $K_n$ satisfy not only $\sum_n K_n^\dagger K_n=\I$ but also $K_n\mathcal{I}K_n^\dagger\subset\mathcal{I}$ for each $K_n$, i.e., each $K_n$ maps an incoherent state to an incoherent state. An incoherent operation is called a strictly incoherent operation or a strictly incoherent channel if $K_n$ also satisfy $K_n^\dagger\mathcal{I}K_n\subset\mathcal{I}$ for each $K_n$ \cite{Winter.Yang2016,Yadin.etal2015}.

A functional $C$ can be taken as a coherence measure if it satisfies the four postulates \cite{Baumgratz.etal2014}:
\\(C1) $C(\rho)\ge 0$, and $C(\rho)=0$ if and only if $\rho\in\mathcal{I}$;
\\(C2) Monotonicity under incoherent operations, $C(\rho)\ge C(\Lambda(\rho))$ if $\Lambda$ is an incoherent operation;
\\(C3) Monotonicity under selective incoherent operations, $C(\rho)\ge \sum_np_nC(\rho_n)$, where $p_n=\Tr(K_n\rho K_n^\dagger)$, $\rho_n=K_n\rho K_n^\dagger/p_n$, and $\Lambda(\rho)=\sum_nK_n\rho K_n^\dagger$ is an incoherent operation;
\\(C4) Non-increasing under mixing of quantum states, i.e., convexity, $\sum_np_nC(\rho_n)\ge C(\sum_np_n\rho_n)$ for any set of states $\{\rho_n\}$ and any probability distribution $\{p_n\}$.

One well-known coherence measure is the relative entropy of coherence $C_r$. It is defined as
\begin{equation}
C_r(\rho)=\min_{\delta\in\mathcal{I}}S(\rho||\delta),
\label{eq:defrec}
\end{equation}
where $S(\rho||\delta)=\Tr\rho(\log\rho-\log\delta)$ is the relative entropy.

With these notions, we can now state our main finding as a theorem.

\begin{theorem}
$C(\rho_t)=C(\rho_0)$ for all coherence measures $C$ if and only if $C_r(\rho_t)=C_r(\rho_0)$, where $\rho_t=\Lambda_t(\rho_0)$ with $\Lambda_t$ being a strictly incoherent channel and $\rho_0$ being an initial state.
\end{theorem}

We only need to prove that $C(\rho_t)=C(\rho_0)$ if $C_r(\rho_t)=C_r(\rho_0)$ in the theorem, since $C_r$ is certainly frozen if all measures are frozen.

First, we show that $S(\Lambda_t(\rho_0)||\Lambda_t(\delta_0))=S(\rho_0||\delta_0)$, where $\delta_0$ is the diagonal part of the density matrix $\rho_0$. By definition, $C_r(\rho)=\min_{\delta\in\mathcal{I}}S(\rho||\delta)$. The minimum is attained if and only if $\delta=\rho_d$, where $\rho_d$ is the diagonal part of $\rho$ \cite{Baumgratz.etal2014}, and then there is
\begin{equation}
 C_r(\rho_0)=S(\rho_0||\delta_0).
\label{eq:cr0}
\end{equation}
By using the contractivity of the relative entropy, i.e., $S(\mathcal{E}(\rho_1)||\mathcal{E}(\rho_2))\le S(\rho_1||\rho_2)$ for any two states $\rho_1$ and $\rho_2$ under a CPTP map $\mathcal{E}$ \cite{Lindblad1975,Uhlmann1977,Nielsen.Chuang2000}, we have
\begin{equation}
S(\Lambda_t(\rho_0)||\Lambda_t(\delta_0))\le S(\rho_0||\delta_0).
  \label{eq:crtge}
\end{equation}
On the other hand, since $\Lambda_t$ is an incoherent channel, there is $\Lambda_t(\delta_0)\in\mathcal{I}$, which further leads to
\begin{equation}
  C_r(\rho_t)=\min_{\delta\in\mathcal{I}}S(\rho_t||\delta)\le S(\rho_t||\Lambda_t(\delta_0)).
  \label{eq:retge}
\end{equation}
Combining Eqs. \eqref{eq:cr0}, \eqref{eq:crtge}, and \eqref{eq:retge}, we obtain the inequality,
\begin{equation}
  C_r(\rho_t)\le S(\rho_t||\Lambda_t(\delta_0))\le C_r(\rho_0).
  \label{eq:crgege}
\end{equation}
In the condition of $C_r(\rho_t)=C_r(\rho_0)$, Eq. (\ref{eq:crgege}) results in
\begin{equation}
  C_r(\rho_t)=S(\rho_t||\Lambda_t(\delta_0)),
  \label{eq:crt}
\end{equation}
and
\begin{equation}
 S(\Lambda_t(\rho_0)||\Lambda_t(\delta_0))=S(\rho_0||\delta_0).
\label{eq:recatt}
\end{equation}
Equation \eqref{eq:crt} indicates that $\Lambda_t(\delta_0)$ is just the diagonal part of the density matrix $\rho_t=\Lambda_t(\rho_0)$, while Eq. \eqref{eq:recatt} shows that the equality for the contractivity of relative entropy in Eq. \eqref{eq:crtge} is attained. Hereafter, we will use $\delta_t$ to denote the diagonal part of the density matrix $\rho_t$ for simplicity. The above discussion implies that $\delta_t=\Lambda_t(\delta_0)$.

Second, we demonstrate that there exists an incoherent operation $R_t$ such that $R_t(\rho_t)=\rho_0$ and $R_t(\delta_t)=\delta_0$. According to the well-known result about the contractivity of relative entropy given in Refs. \cite{Petz1986,Petz2003}, we have that Eq. \eqref{eq:recatt} is valid if and only if there exists a CPTP map $R_t$ such that
\begin{equation}
  R_t(\rho_t)=\rho_0, ~~~ R_t(\delta_t)=\delta_0.
  \label{eq:recovery}
\end{equation}
We therefore only need to prove that this CPTP map is incoherent. In the case that $\delta_t$ is invertible, a CPTP map satisfying Eq. (\ref{eq:recovery}) can be explicitly expressed as \cite{Hayden.etal2004},
\begin{equation}
  R_t(\rho)=\sum_n\delta_0^{\frac{1}{2}}K_n^\dagger(t)\delta_t^{-\frac{1}{2}}\rho\delta_t^{-\frac{1}{2}}K_n(t)\delta_0^{\frac{1}{2}}.
  \label{eq:Petz}
\end{equation}
with the Kraus operators $\tilde{K}_n(t)=\delta_0^{\frac{1}{2}}K_n^\dagger(t)\delta_t^{-\frac{1}{2}}$.  Since $\delta_t^{-\frac{1}{2}}\mathcal{I}\delta_t^{-\frac{1}{2}}\subset\mathcal{I}$, $K_n^\dagger(t)\mathcal{I}K_n(t)\subset\mathcal{I}$, and $\delta_0^{\frac{1}{2}}\mathcal{I}\delta_0^{\frac{1}{2}}\subset\mathcal{I}$, it is easy to verify that $\tilde{K}_n(t)\mathcal{I}\tilde{K}_n^\dagger(t)\subset\mathcal{I}$. Hence, Eq. (\ref{eq:Petz}) defines an incoherent CPTP map satisfying Eq. \eqref{eq:recovery}. In the case that $\delta_t$ is not invertible, instead of Eq. (\ref{eq:Petz}), $R_t$ can be written as
\begin{equation}
  R_t(\rho)=\sum_n\delta_0^{\frac{1}{2}}K_n^\dagger(t)\delta_t^{-\frac{1}{2}}\rho \delta_t^{-\frac{1}{2}}K_n(t)\delta_0^{\frac{1}{2}}+P\rho P,
  \label{eq:Petz2}
\end{equation}
where $P$ is the orthogonal projector onto the eigenspace of $\delta_t$ associated with eigenvalue $0$, and $\delta_t^{-\frac{1}{2}}$ is defined by $(\delta_t^{-\frac{1}{2}})_{ii}=(\delta_t)_{ii}^{-\frac{1}{2}}$ if $(\delta_t)_{ii}\neq 0$, and $(\delta_t^{-\frac{1}{2}})_{ii}=0$ if $(\delta_t)_{ii}=0$. Similarly, we can show that the $R_t$ defined in \eqref{eq:Petz2} is an incoherent CPTP map, and satisfies Eq. \eqref{eq:recovery}.

Third, with the above arguments, it is ready to obtain the conclusion $C(\rho_t)=C(\rho_0)$. By combining the two incoherent operations $\Lambda_t$ and $R_t$, there is
\begin{equation}
  \rho_0\xrightarrow{\Lambda_t}\rho_t\xrightarrow{R_t}\rho_0.
  \label{eq:rho}
\end{equation}
Since all the coherence measures $C$ have the monotonicity of coherence measure under incoherent CPTP map, expressed by the postulate (C2), Eq. (\ref{eq:rho}) results in
\begin{equation}
  C(\rho_0)\ge C(\rho_t)\ge C(\rho_0),
  \label{eq:rhoineq}
\end{equation}
which implies that $C(\rho_t)=C(\rho_0)$. This completes the proof of our theorem.

The theorem means that all measures of coherence are frozen for an initial state in a strictly incoherent channel if and only if the relative entropy of coherence is frozen for the state. It provides an entropy-based criterion for identifying the states with measure-independent freezing of coherence, and is applicable to all strictly incoherent channels.

It is worth noting that all the typical qubit noisy channels \cite{Nielsen.Chuang2000}, such as the bit flip, phase flip, bit-phase flip, depolarizing, phase damping, and amplitude damping channels, belong to this class of channels. It is easy to verify that all the Kraus operators describing these channels satisfy both $K_n(t)\mathcal{I}K_n^\dagger(t)\subset\mathcal{I}$ and $K_n^\dagger(t)\mathcal{I}K_n(t)\subset\mathcal{I}$. Furthermore, if $N$ channels $\Lambda_t^\alpha$ with Kraus operators $K_n^\alpha(t)$, $\alpha=1,2,\ldots,N$, are strictly incoherent channels, then the local channel $\Lambda_t=\Lambda_t^1\otimes\Lambda_t^2\otimes\cdots\otimes\Lambda_t^N$ is also a strictly incoherent channel with its Kraus operators $K_{n_1n_2\ldots n_N}=K_{n_1}^1\otimes K_{n_2}^1\otimes\cdots\otimes K_{n_N}^N$ satisfying $K_{n_1n_2\ldots n_N}\mathcal{I}K_{n_1n_2\ldots n_N}^\dagger\subset\mathcal{I}$ and $K_{n_1n_2\ldots n_N}^\dagger\mathcal{I}K_{n_1n_2\ldots n_N}\subset\mathcal{I}$. Note that here $\Lambda_t^1,\Lambda_t^2,\ldots,\Lambda_t^N$ need not be identical, i.e., they may be different noisy channels. In fact, $K_n(t)\mathcal{I}K_n^\dagger(t)\subset\mathcal{I}$ means that there is at most one nonzero entry in each column of $K_n$ \cite{Yao.etal2015}, while similarly $K_n^\dagger(t)\mathcal{I}K_n(t)\subset\mathcal{I}$ means that there is at most one nonzero entry in each row of $K_n$. Therefore, a channel is a strictly incoherent channel if and only if at most one nonzero entry appears in each row and each column of its Kraus operators with respect to the fixed reference basis. This provides a simple approach to identify strictly incoherent channels, by which it is very easy to confirm that all the local channels consisting of strictly incoherent channels are strictly incoherent channels. Hence our theorem is applicable to all local channels consisting of the typical qubit noisy channels.

Our theorem can help to effectively identify the states with measure-independent freezing of coherence in a strictly incoherent channel. All the states can be obtained only by solving the equation $C_r(\Lambda_t(\rho_0))=C_r(\rho_0)$, although it may be difficult to solve analytically the equation to obtain the whole solutions since the calculation of entropy is complicated. However, in general, it is unnecessary to obtain all the solutions. In quantum information processing, researchers are usually interested only in some special states, such as the Bell states, GHZ states, and some other special families of states. In this case, we only need to examine the desired states, to which our theorem is quite useful.

As an example, we now apply our theorem to local bit flip channels to show the measure-independent freezing phenomenon of coherence. Consider an $N$-qubit system undergoing a local bit flip channel $\Lambda_t=\Lambda_{q_1}^1\otimes\dots\otimes\Lambda_{q_N}^N$, where $\Lambda_{q_\alpha}^\alpha(\rho)=K_0^\alpha\rho K_0^{\alpha\dagger}+K_1^\alpha\rho K_1^{\alpha\dagger}$ is the bit flip operation on the $\alpha$-th qubit with $K_0^\alpha=\sqrt{1-q_\alpha}\I$ and $K_1^\alpha=\sqrt{q_\alpha}\sigma_1$, and $q_1,\dots,q_N$ are parameters dependent on time $t$. Here $\sigma_1$ is the Pauli-X operator.

We first examine a family of pure states, defined by
\begin{equation}
  \ket{\varphi_{l_1l_2\ldots l_N}^\pm}=\frac{\ket{l_1l_2\ldots l_N}\pm\ket{\bar{l}_1\bar{l}_2\ldots\bar{l}_N}}{\sqrt{2}},
  \label{eq:GHZ_l_pm}
\end{equation}
 where $l_1=0$, $l_{i\neq1}=0,1,$ and $\bar{l}_i=\NOT(l_i)=1-l_i$. These states are widely used in quantum information processing. For instance, at $N=2$, $\ket{\varphi_{00}^\pm}=\frac{\ket{00}\pm\ket{11}}{\sqrt{2}}$ and $\ket{\varphi_{01}^\pm}=\frac{\ket{01}\pm\ket{10}}{\sqrt{2}}$ are just the Bell states, and at $N\ge 3$, $\ket{\varphi_{00\ldots 0}^+}=\frac{\ket{0}^{\otimes N}+\ket{1}^{\otimes N}}{\sqrt{2}}$ are just the $N$-qubit GHZ states. We will show that all coherence measures for each of the states in Eq. (\ref{eq:GHZ_l_pm}) are frozen.

 Hereafter, we use $l$ ($\bar{l}$) to denote the sequence $l_1 l_2 \ldots l_N$ ($\bar{l}_1\bar{l}_2\ldots\bar{l}_N$) for simplicity. The expression in Eq. (\ref{eq:GHZ_l_pm}) can then be simply written as $\ket{\varphi_l^\pm}=\frac{\ket{l}\pm\ket{\bar{l}}}{\sqrt{2}}$. According to our theorem, we only need to show that the relative entropy $C_r(\rho_{t,l}^{\pm})$ are constants, where $\rho_{t,l}^{\pm}=\Lambda_t(\rho_{0,l}^{\pm})$ with $\rho_{0,l}^{\pm}=\ket{\varphi_{l}^\pm}\bra{\varphi_{l}^\pm}$ being the initial states.

 By detail calculations, we obtain
\begin{equation}
  \rho_{t,l}^{\pm}=\sum_{l'}p_{t,l'l}\ket{\varphi_{l'}^\pm}\bra{\varphi_{l'}^\pm},
  \notag
\end{equation}
where
\begin{equation}
 \begin{aligned}
    p_{t,l'l}=&\prod_{\substack{1\le i\le N}}\Big(q_i+(1-2q_i)\delta_{l_i'l_i}\Big)\\
    &+\prod_{\substack{1\le i\le N}}\Big(1-q_i-(1-2q_i)\delta_{l_i'l_i}\Big).
    \notag
 \end{aligned}
\end{equation}
The $2^N$ eigenvectors of $\rho_{t,l}^\pm$ can be taken as $\ket{\varphi_{l'}^+}$ and $\ket{\varphi_{l'}^-}$, which satisfy
\begin{equation}
  \begin{aligned}
    &\rho_{t,l}^\pm\ket{\varphi_{l'}^+}=\frac{p_{t,l'l}\pm p_{t,l'l}}{2} \ket{\varphi_{l'}^+}, ~~\\ &\rho_{t,l}^\pm\ket{\varphi_{l'}^-}=\frac{p_{t,l'l}\mp p_{t,l'l}}{2}\ket{\varphi_{l'}^-},
  \end{aligned}
  \notag
\end{equation}
and the diagonal part of $\rho_{t,l}^\pm$ is
\begin{equation}
  \delta_{t,l}=\sum_{l'}\Big(\frac{1}{2}p_{t,l'l}\ket{l'}\bra{l'}+\frac{1}{2}p_{t,l'l}\ket{\bar{l}'}\bra{\bar{l}'}\Big).
  \notag
\end{equation}
With the aid of the above expressions, we can calculate the relative entropy of coherence, and obtain
\begin{equation}
  \begin{aligned}
    &C_r(\rho_{t,l}^\pm)=S(\rho_{t,l}^\pm||\delta_{t,l})=S(\delta_{t,l})-S(\rho_{t,l}^\pm)\\
     =&-\sum_{l'}\Big(p_{t,l'l}\log(p_{t,l'l})+p_{t,l'l}\log\frac{1}{2}\Big)+\sum_{l'}p_{t,l'l}\log p_{t,l'l}\\
     =&\sum_{l'}p_{t,l'l}=1.
  \end{aligned}
  \label{eq:C_r_example_1}
\end{equation}
Equation (\ref{eq:C_r_example_1}) shows that the relative entropy of coherence for each state $\rho_{t,l}^\pm$ is constant, and therefore all coherence measures manifest freezing forever for the $N$-qubit system initially in the states expressed by Eq. (\ref{eq:GHZ_l_pm}) undergoing local bit flip channels. The measure-independent freezing occurs in this case.

We now extend our discussion to a family of mixed states, defined by
\begin{equation}
  \rho_0=\sum_lp_l\left(p\ket{\varphi_l^+}\bra{\varphi_l^+}+(1-p)\ket{\varphi_l^-}\bra{\varphi_l^-}\right),
  \label{eq:GHZ_mixed}
\end{equation}
where $0\le p\le 1$, and $\{p_l\}$ is any probability distribution. Again, $\ket{\varphi_l^\pm}=\frac{\ket{l}\pm\ket{\bar{l}}}{\sqrt{2}}$ are the brief expression of the pure states defined in Eq. (\ref{eq:GHZ_l_pm}).

For the local bit flip channel $\Lambda_t=\Lambda_{q_1}^1\otimes\dots\otimes\Lambda_{q_N}^N$, we have
\begin{equation}
\rho_t=\Lambda_t(\rho_0)=\sum_lp_{t,l}\left(p\ket{\varphi_l^+}\bra{\varphi_l^+}+(1-p)\ket{\varphi_l^-}\bra{\varphi_l^-}\right),
\notag
\end{equation}
where
\begin{equation}
  \begin{aligned}
  p_{t,l}=&\sum_{l'}p_{l'}\Big(\prod_{\substack{1\le i\le N}}(q_i+(1-2q_i)\delta_{l_i'l_i})\\
    &+\prod_{\substack{1\le i\le N}}(1-q_i-(1-2q_i)\delta_{l_i'l_i})\Big).
  \end{aligned}
  \notag
\end{equation}
The $2^N$ eigenvectors of $\rho_t$ can be taken as $\ket{\varphi_l^+}$ and $\ket{\varphi_l^-}$, which satisfy
\begin{equation}
  \rho_t\ket{\varphi_l^+}=p_{t,l}p\ket{\varphi_l^+}, \quad \rho_t\ket{\varphi_l^-}=p_{t,l}(1-p)\ket{\varphi_l^-},
  \notag
\end{equation}
and the diagonal part of $\rho_t$ is
\begin{equation}
  \delta_t=\sum_l\Big(\frac{1}{2}p_{t,l}\ket{l}\bra{l}+\frac{1}{2}p_{t,l}\ket{\bar{l}}\bra{\bar{l}}\Big).
  \notag
\end{equation}
We can then obtain the relative entropy of coherence,
\begin{equation}
  \begin{aligned}
     &C_r(\rho_t)=S(\delta_t)-S(\rho_t)\\
    =&-\sum_l\Big(p_{t,l}\log p_{t,l}+p_{t,l}\log\frac{1}{2}\Big)\\
    +&\sum_l\Big(p_{t,l}\log p_{t,l}+p_{t,l}(p\log p+(1-p)\log(1-p))\Big)\\
    =&1-H(p),\\
  \end{aligned}
  \label{eq:C_r_example_2}
\end{equation}
where $H(p)=-p\log p-(1-p)\log(1-p)$, being the binary Shannon entropy. Equation (\ref{eq:C_r_example_2}) shows that $C_r(\rho_t)$ is a constant, which implies that all coherence measures manifest freezing forever for the $N$-qubit system initially in the states expressed by Eq. (\ref{eq:GHZ_mixed}) undergoing local bit flip channels, i.e., the measure-independent freezing occurs.

Specially, if we take $N$ as even numbers and let $p=\frac{1+c_1}{2}$ and $p_l=\frac{1+(-1)^{w(l)}c_3}{2^{N-1}}$ in our example,
where $-1\le c_1,c_3\le 1$ are two real number and $w(l)$ is the Hamming weight of $\ket{l}$, then Eq. (\ref{eq:GHZ_mixed}) gives the states discussed in Ref. \cite{Bromley.etal2015}, where the authors found that the coherence measures based on the bona fide distance are frozen in the local identical bit flip channel $\Lambda_t=\Lambda_q^{\otimes N}$. Here, our example implies that in this case all coherence measures, not limited to the bona fide coherence measures, are frozen.

In conclusion, we have proved the theorem that all measures of coherence are frozen for an initial state in a strictly incoherent channel if and only if the relative entropy of coherence is frozen for the state. Our finding reveals the existence of measure-independent freezing of coherence, and more importantly, provides an entropy-based dynamical condition in which the coherence of an open quantum system is totally unaffected by noise.

Our theorem is applicable to all strictly incoherent channels, such as the typical channels including the bit flip, phase flip, bit-phase flip, depolarizing, phase damping, amplitude damping channels, and all the multiqubit local noisy channels consisting of these typical qubit channels. As an example, we have applied the theorem to local bit flip channels, and shown that there are a number of states including the Bell states, the GHZ states, other pure states, and a family of mixed states, of which all coherence measures are frozen.

In passing, we would like to add that the relative entropy of coherence plays a crucial role in the theorem. We do not find other coherence measures which can take the place of the relative entropy in the theorem.

\begin{acknowledgments}
This work was supported by NSF China through Grant No. 11575101 and the National Basic Research Program of China through Grant No. 2015CB921004. D.M.T. acknowledges support from the Taishan Scholarship Project of Shandong Province.
\end{acknowledgments}


%

\end{document}